\documentclass[11pt]{revtex4}
\usepackage{amssymb,epsf}
\usepackage{epsfig}
\usepackage{graphicx, latexsym, amssymb, amscd, psfrag}
\usepackage{latexsym}
\begin{document}

\title{Double field domain walls with explicit symmetry breaking}
\author{ Nematollah Riazi$^{1}$\footnote{email: riazi@physics.susc.ac.ir, on leave from the Physics Department of Shiraz
University.}, Marzieh Peyravi$^{2}$\footnote{email:
Marziyeh.Peyravi@stu-mail.um.ac.ir} and Shahram
Abbassi$^{2}$\footnote{email: abbassi@ipm.ir} }
\address{$^1$. Physics Department, Shahid Beheshti University, Evin, Tehran 19839, Iran\\
$^2$. Department of Physics, School of Sciences, Ferdowsi
University of Mashhad, Mashhad 91775-1436, Iran.}

\begin{abstract}
We study the dynamics of domain walls in a double-field model in
which the U(1) symmetry is broken both spontaneously and
explicitly. The global U(1) symmetry of the system is restored
when the symmetry breaking parameter $\epsilon$ is set to zero.
Two pairs of degenerate kinks exist in the model with are related
to each other by a $Z_2$ transformation.  We first calculate the
single domain wall solutions and then investigate collision
processes. These include simple scattering, pair annihilation,
pair capture, and other interesting processes. The possibility of
the domain wall being punctured by a string is also investigated.

\end{abstract}

\maketitle
\section{Introduction\label{intro}}
Domain walls are topological solitons which appear in certain
nonlinear field equations having disconnected degenerate
vacuua\cite{man}. These interesting objects occur in different
systems, including phase transitions in the early
universe\cite{dom}, magnetism\cite{mag}, optics\cite{opt}, and
brane world scenarios\cite{rub}. Once formed, domain walls can
bend, collide, and annihilate each other\cite{dom}. When viewed as
one-dimensional, localized objects, domain walls are usually
called kinks. This is why we use the "domain wall" and "kink"
terms interchangeably throughout this paper.

Apart from gravitational interactions, domain walls and kinks
interact with each other via short range forces and collide
without losing their identities \cite{0,1,2,3,4,5}.  Like other
topological solitons, domain walls are stable, due to the boundary
conditions at spatial infinity from the wall. Their existence,
therefore, is essentially dependent on the presence of degenerate
vacua \cite{1,6}. In the case of domain walls in three spatial
dimensions, the curvature of the wall also leads to acceleration.
This acceleration can lead to the emission of scalar and
gravitational radiation\cite{dom}.

Topology provides an elegant way of classifying domain walls in
various sectors according to the mappings between the degenerate
vacua of the field and the points at spatial infinity\cite{man}.
For the Sine-Gordon (SG) system in $1+1$ dimensions, these
mappings are between $\phi=2n\pi$, $n\in\mathbb{Z}$ and
$x=\pm\infty$, which correspond to kinks and antikinks of the SG
system. More complicated mappings occur for solitons in higher
dimensions \cite{2}.

Coupled systems of scalar fields with soliton solutions have found
interesting applications in double-strand, long molecules like the
DNA molecules \cite{dna,yak,cu,pey}, bi-dimensional QCD
\cite{qcd}, and hybrid (double-field) inflationary model in
cosmology \cite{linde}. Analytical and numerical properties of
such models are investigated by many authors, including Bazeia et
al \cite{ba} and Riazi et al \cite{1,riman}. Inspired by the
coupled systems introduced in \cite{1,7a}, we investigate a new
coupled system of two real scalar fields. The present model may be
used as a tentative model of a double-field inflation. In the
present paper, we are interested in the domain wall interactions
within this model.

The field potential we start with reads
\begin{equation}\label{a}
V(\phi,\psi)=(\phi^2+\psi^2-1)^2+\frac{1}{2}\lambda\psi^2,
\end{equation}
in which $\phi$ and $\psi$ are real scalar fields, and $\lambda$
is a constant controlling the explicit symmetry breaking. This
potential is similar, but not the same as that of the hybrid
inflationary model\cite{linde}. Note that the potential along the
$\phi$ axis has always two degenerate vacua at $\phi=\pm 1$, while
the potential along the $\psi$ axis has minima at $\psi=\pm 1$
only if $\lambda < 4$. For $\lambda \ge 4$, the potential has only
one minimum at $\psi =0$ along this axis (besides the two absolute
minima). The two minima at $\psi=\pm 1$ are in fact saddle points
for $\lambda < 4$.

In the hybrid inflationary model, there are two scalar fields, one
playing the role of rapidly decaying (water-fall) field, triggered
by another (inflationary) scalar field\cite{linde}. Depending on
the choice of the Lagrangian density, the model may lead to the
formation of domain walls. In what follows, we show that the
potential (1) leads to the formation of domain walls and in
subsequent sections, we investigate how they interact with each
other.

\section{Preliminaries}\label{sec2}
The Lagrangian density of the system is given by:
\begin{equation}\label{a}
{\cal L}=
\frac{1}{2}\partial^{\mu}\phi\partial_{\mu}\phi+\frac{1}{2}\partial^{\mu}\psi\partial_{\mu}\psi
-[(\phi^2+\psi^2-1)^2+\frac{1}{2}\lambda\psi^2],
\end{equation}
in which $\lambda$ is the $U(1)$ explicit symmetry breaking
parameter. One can write this Lagrangian in terms of the complex
scalar field $\Phi$, where
\begin{equation}\label{b}
\Phi = \phi+i\psi,
\end{equation}
in terms of which the Lagrangian density of the system reads
\begin{equation}\label{c}
{\cal L}= \frac{1}{2}\partial^{\mu}\Phi^\dag\partial_{\mu}\Phi-
(\Phi^\dag\Phi-1)^2+\frac{1}{2}\lambda(Im \Phi)^2 .
\end{equation}
From either of these two forms of the Lagrangian density, the
following field equations for $\phi$ and $\psi$ are obtained:
\begin{equation}\label{d}
\Box\phi=-4\phi(\phi^2+\psi^2-1);
\end{equation}
and
\begin{equation}\label{e}
\Box\psi=-4\psi(\phi^2+\psi^2-1)+\lambda\psi.
\end{equation}
It is obvious that  if $\lambda=0$, the Lagrangian density
(\ref{c}) is both Lorentz invariant and also invariant under a
global $U(1)$ symmetry
\begin{equation}
\Phi\rightarrow \Phi'=e^{i\theta}\Phi,
\end{equation}
or
\[
\phi'=\phi \cos \theta -\psi \sin \theta,
\]
\begin{equation}
\psi'=\phi \sin \theta +\psi \cos \theta .
\end{equation}
 The corresponding energy-momentum tensor\cite{7,8} of the system is:
\begin{equation}\label{f}
T_{\mu\nu} =
\partial_{\mu}\phi\partial_{\nu}\phi+\partial_{\mu}\psi\partial_{\nu}\psi
- g_{\mu\nu}{\cal L};
\end{equation}
which satisfies the conservation law
\begin{equation}\label{g}
\partial_{\mu}T^{\mu\nu} =0.
\end{equation}
In Equation (\ref{f}),  $g_{\mu\nu}=diag(1,-1,-1,-1)$ is the
metric of the $(3+1)$-dimensional spacetime for $\phi$ and $\psi$
functions of $x$ and $t$. The Hamiltonian (energy) density is
obtained from Eq.(\ref{f}) according to
\begin{equation}\label{ff}
\mathcal{H}=T^{00}=\frac{1}{2}\left(\frac{\partial \phi}{\partial
t}\right)^2+\frac{1}{2}\left(\frac{\partial \psi}{\partial
t}\right)^2+\frac{1}{2}\left(\frac{\partial \phi}{\partial
x}\right)^2+\frac{1}{2}\left(\frac{\partial \psi}{\partial
x}\right)^2+V(\phi,\psi).
\end{equation}
In order to derive the domain wall solutions, one has to reduce
the system to an effectively 1+1 dimensional spacetime by assuming
the fields depending on one space and one time coordinate. In
fact, a domain wall is nothing but a kink placed inside a 3D
space\cite{dom}. Like other systems bearing topological solitons,
the present system also has the following topological current:
\begin{equation}\label{g}
J^{\mu}=\frac{1}{2}\epsilon^{\mu\nu}\partial_{\nu}\phi
\end{equation}
which is locally conserved:
\begin{equation}\label{h}
\partial_{\mu}J^{\mu}=0.
\end{equation}
 The corresponding topological charge is given by:
\begin{equation}\label{i}
Q=\int_{-\infty}^{+\infty}J^{0}dx=\frac{1}{2}[\phi(+\infty)-\phi(-\infty)].
\end{equation}
Note that since the vacua of the system reside at
$(\phi,\psi)=(\pm 1, 0)$, only the $\phi$-field is responsible for
the topological charge.

According to the Goldstone theorem, if a continuous global symmetry
is broken spontaneously, there appears a massless (Goldstone)
particle for each broken group parameter\cite{nam,gol}. However, in
the case of the Lagrangian density (\ref{c}), in addition to the
spontaneous breaking of the U(1) symmetry for $\lambda=0$, the
symmetry is broken explicitly by the $\lambda$-term. If we expand
the potential around either of the vacua $(\phi=\pm 1,\psi=0)$,
there appears the following mass terms:
\begin{equation}
V(\chi , \psi)\simeq \frac{1}{2}m_\chi \chi^2 +\frac{1}{2}m_\psi
\psi^2,
\end{equation}
where
\begin{equation}
m_\chi\equiv \frac{\partial^2 V}{\partial \phi^2}|_{(\phi=\pm
1,\psi=0)}=8,
\end{equation}
and
\begin{equation}
m_\psi\equiv \frac{\partial^2 V}{\partial \psi^2}|_{(\phi=\pm
1,\psi=0)}=\lambda,
\end{equation}
where $\chi\equiv \phi-1$. It is seen that due to the explicit
symmetry breaking term, the massless Goldstone boson ($\psi$)
which is normally massless, has acquired a mass ($\lambda$).

Let us consider the $U(1)$-symmetric case $\lambda=0$. The
potential reduces to the well-known complex $\varphi^{4}$ model
and we have the global $U(1)$ symmetry
\begin{eqnarray}\label{ee}
\Phi\longrightarrow {\Phi}'&=& e^{i\theta}\Phi,\nonumber\\
{{\cal L}}'&=&{\cal L}.
\end{eqnarray}
This symmetry leads to the following conserved current and charge,
as deduced from the celebrated Noether's theorem\cite{8}
\begin{equation}\label{h}
J_N^{\mu}=
i(\Phi^{\ast}\partial^{\mu}\Phi-\Phi\partial^{\mu}\Phi^{\ast}).
\end{equation}
\begin{equation}\label{i}
Q_{N}=\int J_N^{0}dx.
\end{equation}
Writing the complex scalar field $\Phi$ in the form
$\Phi=Re^{i\xi}$, the current (\ref{h}) and the charge (\ref{i})
take the following simple forms:
\begin{equation}
J^\mu_N=2R^2\partial^\mu \xi, \end{equation} and
\begin{equation}
Q_N=2\int^{+\infty}_{-\infty}R^2\partial^0 \xi dx.
\end{equation}
It is obvious that the Noether charge vanishes for all static
solutions, including the static kinks and antikinks to be
introduced shortly. For a time-varying field like
$\Phi=R(x)e^{i\omega t}$, however, we have the non-vanishing
Noether charge $Q_N=2\omega \int R^2(x)dx$.

By using the dynamical equations, it can be easily shown that the
$U(1)$ current is partially conserved and we have
\begin{equation}
\partial_\mu J^\mu_N =2\lambda \phi\psi,
\end{equation}
which is proportional to $\lambda$, like the situation arising in
PCAC (partially conserved axial current). In PCAC, the explicit
symmetry breaking term in the Lagrangian is usually assumed to be
linear in $\psi$\cite{8}.

As we shall see in the next section, the symmetry of the system
under  $\phi\leftrightarrow \phi$ and $\psi\leftrightarrow -\psi$
leads to the appearance of two similar domain walls with the same
energy per unit surface. The existence of a $Z_2$ symmetry
breaking term in the potential (e.g. $\kappa \psi^n$, $n$=odd)
lifts this degeneracy and makes punctured domain walls possible
(see section \ref{punsec}).

\section{Domain wall solutions}\label{sec3}

Kink solutions in more than one spatial dimension form sheet-like
structures called domain walls. When placed in 3D Euclidean space,
the domain wall may be represented by a $(xy)$ planar
concentration of energy with the energy density along the $z$-axis
highly peaked at a certain $z$.

\begin{figure}[h]
\epsfxsize=10cm\centerline{\epsfbox{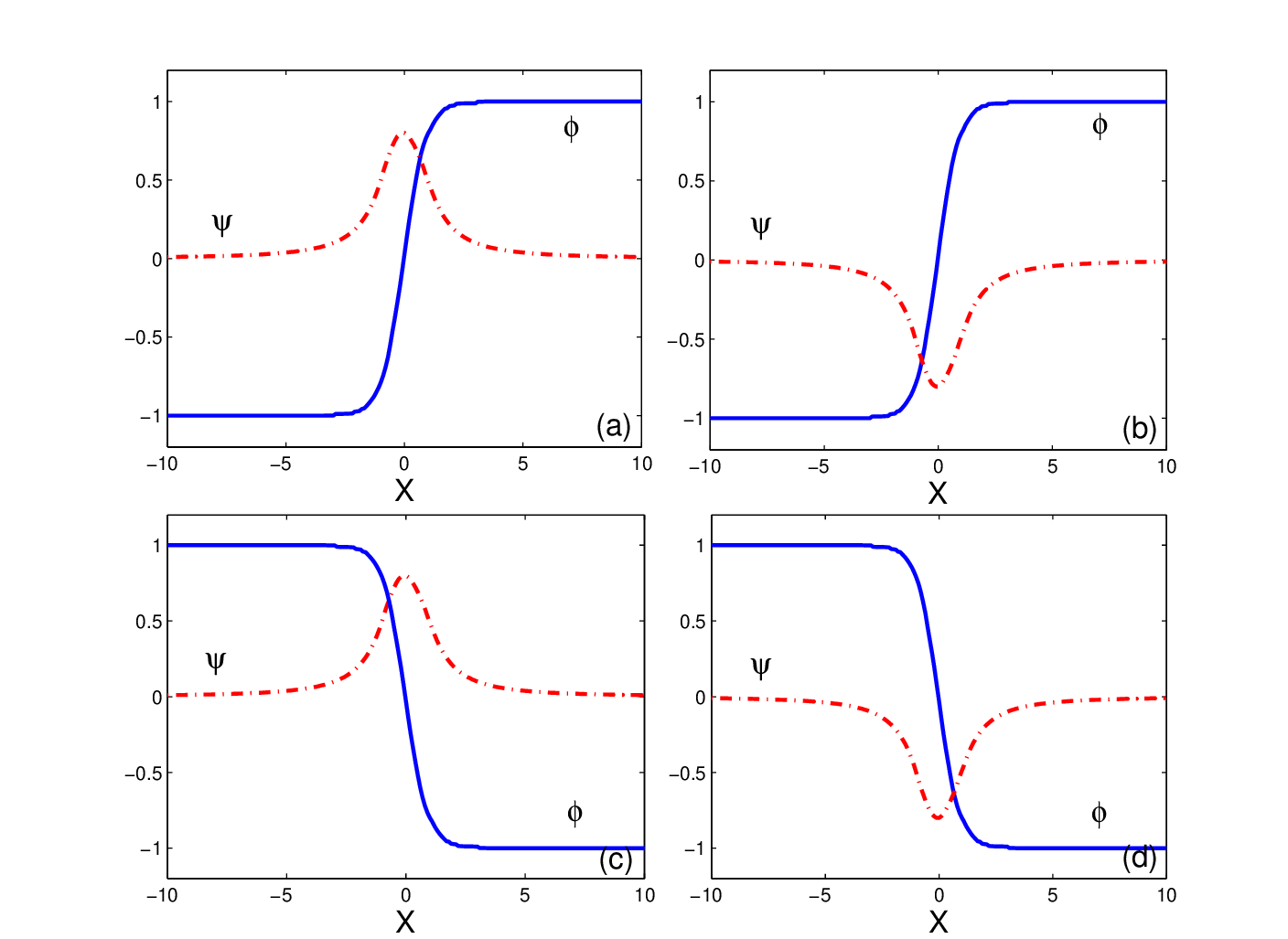}} \caption{Minimum
energy,  static solutions. The solid curves represent $\phi$ and
the dash-dotted curves are for $\psi$.\label{3}}
\end{figure}

For some systems (like the sine-Gordon or the $\phi^4$ systems)
the kink (domain wall) solution can be found analytically. For
many others, including the system under consideration analytical
solutions cannot be found and one must use numerical methods. In
order to find the static kink and antikink solutions which play
the role of domain walls with opposite topological charges, we
have employed the following numerical procedure. The algorithm
starts with an approximate solution which is the exact solution of
the $U(1)$-symmetric system ($\lambda=0$). The solution is then
varied via small changes in the field values, and the total energy
per unit area of the domain wall, as given by
\begin{equation}
E=\int Hdx
\end{equation}
is calculated at each step. The small changes in the field values
is accepted if the total energy is reduced in each step, otherwise
it is rejected. This procedure is iterated repeatedly, until the
program reaches a minimum energy configuration. Domain wall (kink
and anti-kink)) solutions obtained in this way are shown in
Figures \ref{3}.

It is well known that in many nonlinear equations bearing
topological solitons, static solutions satisfy the so-called
Bogomolny condition. This condition puts a lower bound on the
total energy of the system which is proportional to the
topological charge\cite{bog,man}. Multiplying equation (\ref{d})
by $\phi '$ and adding it to equation (\ref{e}) multiplied by
$\psi '$, we obtain
\begin{equation}
\phi ' \phi '' +\psi '\psi '' =\left(\frac{1}{2}(\phi ')^2
+\frac{1}{2}(\psi ')^2\right)'=\frac{\partial V}{\partial
\phi}\phi ' + \frac{\partial V}{\partial \psi}\psi '
=\frac{dV}{dx}.
\end{equation}
Here, prime means derivative with respect to $x$. We thus obtain
the following first integral of the static field equations:
\begin{equation}\label{firstint}
\frac{1}{2}(\phi ')^2+\frac{1}{2}(\psi ')^2-V(\phi,\psi)=C,
\end{equation}
where $C$ is a constant of integration. For localized solutions,
this constant should be zero, since the fields rest on their
vacuum values at $x\rightarrow \pm \infty$. It is seen that there
are two types of kinks and antikinks which are related to each
other by the field transformations $\phi\leftrightarrow \phi$ and
$\psi\leftrightarrow -\psi$, or a simple parity operation.


For low energy density walls, one can use Newtonian formulation to
find the gravitational field via the Poisson equation:
\begin{equation}
\nabla^2 \Phi =4\pi G\rho.
\end{equation}
where $\Phi$ is the gravitational potential and $\rho$ is the
domain wall equivalent mass density, given by $\rho=H/c^2$, $H$
being the Hamiltonian density $H=T^0_0$. Using the Gauss's law for
a cylindrical volume bisected by the domain wall, one finds
\begin{equation}
{\bf g}=-2\pi G\sigma {\bf k},
\end{equation}
for large distances from the wall (compared to the thickness of
the wall). In this equation,  ${\bf g}$ is the gravitational field
strength vector and ${\bf k}$ is the outward unit vector
perpendicular to the domain wall, and $\sigma$ is the domain wall
mass density.

When $g$ is comparable to $c^2/z$ where $c$ is the velocity of
light and $z$ is the distance from the wall, Newtonian theory
breaks down and one has to use general theory of relativity.

The gravitational effects of a planar domain wall in the thin-wall
limit can be found in \cite{167}, \cite{169}.  In the thin-wall
limit one assumes that the domain wall is infinitely thin so only
the vacuum Einstein equations need to be solved on either side of
the wall. By matching the vacuum solutions on the two sides of the
wall (i.e. implementing the junction conditions) which is
facilitated by using the Gauss-Codazzi formalism one can obtain
the appropriate metric \cite{80}.

In this case, we have to use the following metric:
\begin{equation}
ds^2=-f(z)dt^2+h(z)(dx^2+dy^2)+dz^2,
\end{equation}
where $f(z)$ and $h(z)$ are unknown functions. The Einstein
equations with the $\phi$ and $\psi$ fields as sources read
\begin{equation}
G^\mu_\nu=8\pi G T^\mu_\nu=8\pi G \left[ \partial^\mu \phi
\partial_\nu \phi+\partial^\mu \psi
\partial_\nu \psi-\delta^\mu_\nu(\frac{1}{2}\partial^\alpha\phi
\partial_\alpha \phi +\frac{1}{2}\partial^\alpha\psi \partial_\alpha \psi-V(\phi,\psi)) \right].
\end{equation}
We also have the following field equations for $\phi$ and $\psi$
\begin{equation}
\nabla^\mu\nabla_\mu \phi +\frac{\partial V}{\partial \phi}=0,
\end{equation}
and
\begin{equation}
\nabla^\mu\nabla_\mu \psi +\frac{\partial V}{\partial \psi}=0,
\end{equation}
where $\nabla_\mu$ is the covariant derivative. Some single field
gravitational domain wall solutions have been discussed  in
\cite{181} for the case when $16\pi G <\phi>^2<<1$ where $<\phi
>$ is the vacuum expectation value of the field $\phi$. According
to \cite{181}, no essentially different general relativistic
effect is reported. However, when $16\pi G <\phi>^2>1$, new
effects are observed \cite{101,102,170}. It should be noted that
at large inter-wall distances scalar field interactions can be
neglected, since they are short-range, while at short distances
the reverse is true and the nonlinear scalar field interactions
take over. In the next section, we have done our numerical
calculations in the limit where the gravitational effects can be
ignored.

\section{Domain wall collisions}
Flat domain walls are essentially kink solitons. As in other
kink-bearing systems, an important question is the form of the
inter-kink potential and the behavior of the solitons in collisions
with each other. Kinks and antikinks of different nonlinear systems
behave differently in collisions. In most cases, the following
situations arise: 1) A pair of kinks or antikinks which have the
same topological charge repel each other. They retain their original
shape after the collision. In the sine-Gordon system which is
integrable, the pair retain their original speeds after the
collision. In non-integrable systems like the $\phi^4$ system, part
of the energy is converted into small amplitude waves which are
radiated away and the final speed of the solitons is less than their
initial speed\cite{raja}. 2) In the sine-Gordon system, the
collision between a kink and an antikink does not lead to their
destruction and the pair retain their initial speeds after the
collision. The force between the pair is
velocity-dependent\cite{righ}. It is attractive at relatively large
distances and repulsive at short distances. In non-integrable
systems like $\phi^4$, $\phi^6$ or double-sine-Gordon system, the
collision process between a kink and an antikink is more complicated
and interesting phenomena happen\cite{good,hosein}. For example, the
pair annihilate each other when their relative velocity is smaller
than a first threshold $v_1$. For velocities larger than $v_1$ and
smaller than a second threshold $v_2$, there appear scattering
windows in which the pair leave the interaction region with a speed
smaller than their initial speed. Some small amplitude waves are
radiated away in this process. Velocities larger than $v_2$ lead to
the scattering of the pair and emission of radiation. 3) In the
sine-Gordon system, there is a bound state (breather) solution in
which a kink and an antikink oscillate around the center of mass of
the system indefinitely. Breather solutions in non-integrable
systems like $\phi^4$ are unstable and lead to the annihilation of
the pair after transient oscillations.

In this section, we look for the above possibilities in the system
under investigation. Since analytical calculations are not
possible here, we employ the modified finite-difference method
described in \cite{1}.

\begin{figure}[h]
\epsfxsize=10cm \centerline{\epsfbox{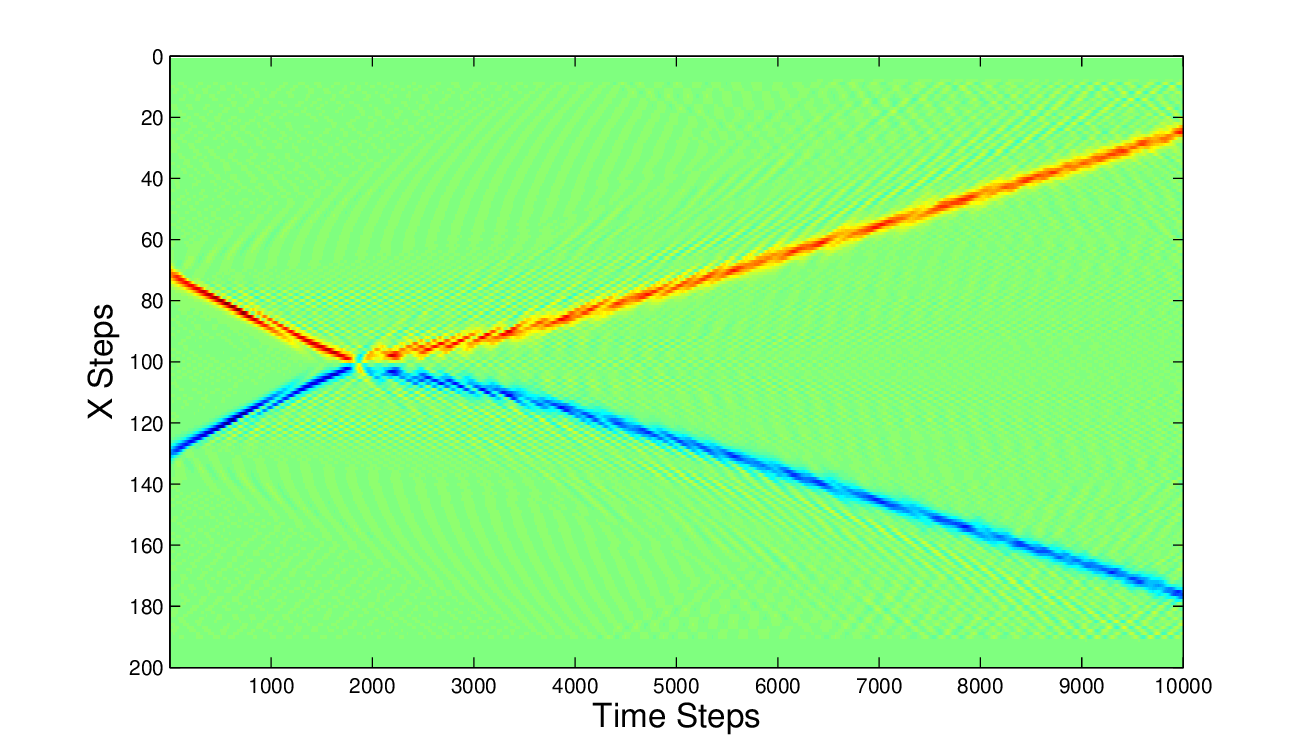}} \caption{Domain
wall (kink-antikink) simple scattering at $v=0.6$. Topological
charge density is plotted on the ($x,t$) plane. Red color
indicates positive and blue indicates negative topological charge.
It is assumed that the collision is side-by-side and the
gravitational effects are ignored. \label{simple}}
\end{figure}

\begin{figure}[h]
\epsfxsize=10cm \centerline{\epsfbox{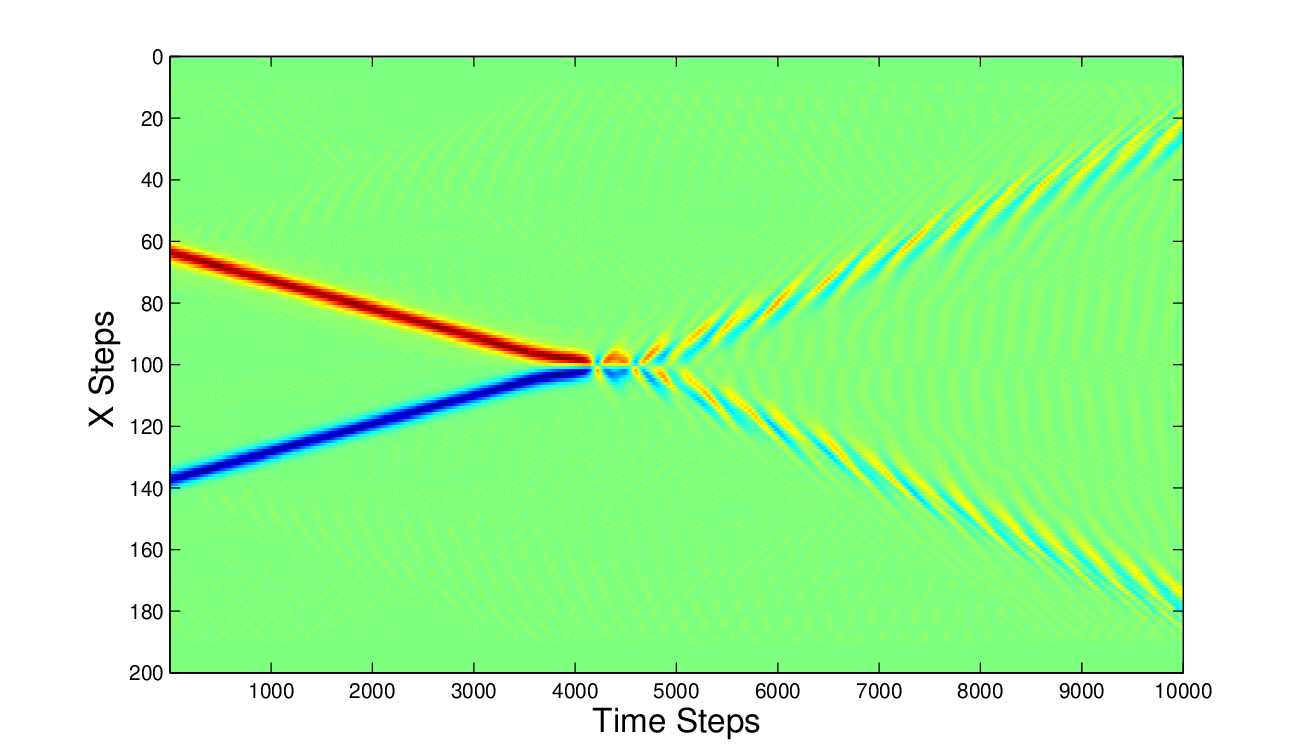}}
\caption{Annihilation of a kink-antikink pair into a pair of
neutral wave packets at $v=0.36$. Red color indicates positive and
blue indicates negative topological charge.\label{gamma}}
\end{figure}

\begin{figure}[h]
\epsfxsize=10cm \centerline{\epsfbox{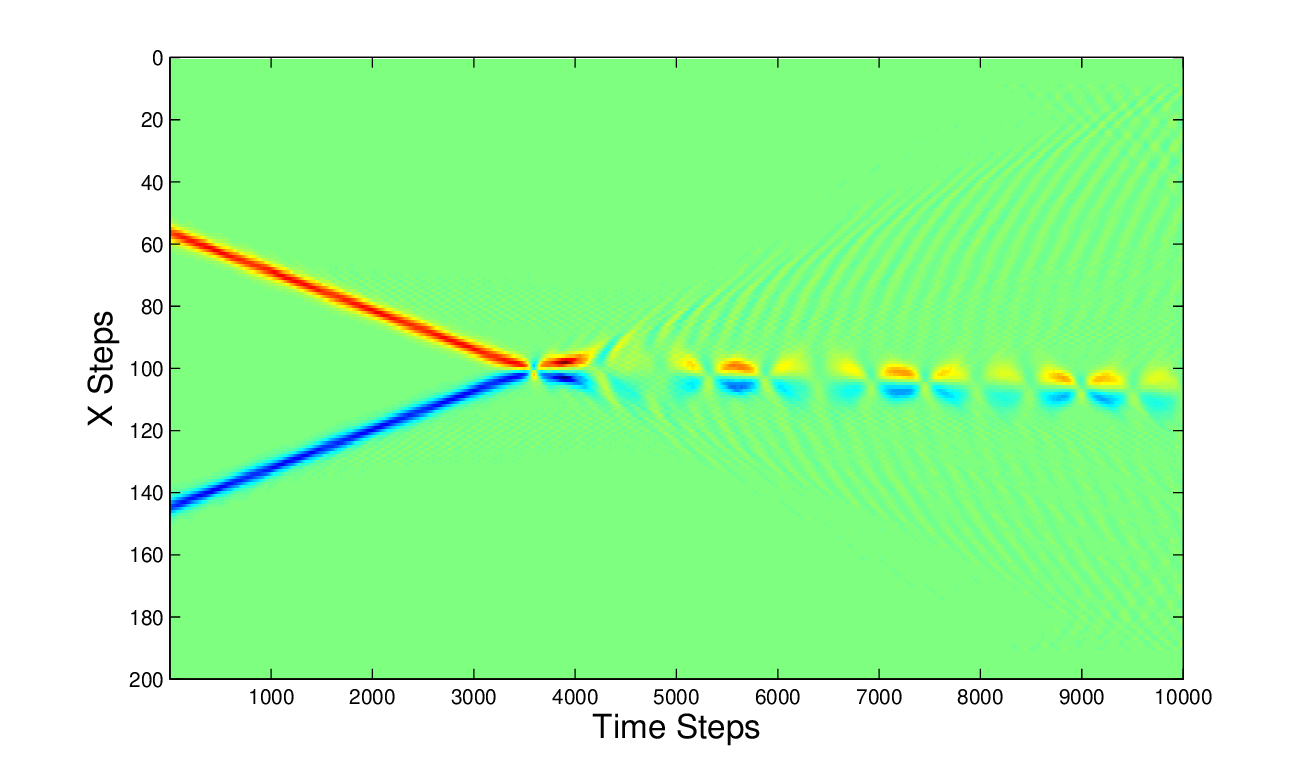}}
\caption{Formation of a soliton molecule via kink-antikink capture
at $v=0.5$. Note that the surplus energy is radiated away. Red
color indicates positive and blue indicates negative topological
charge. This system is different from a breather, since in a
breather the positively and negatively charged kinks periodically
interchange their position.\label{molecule}}
\end{figure}

\begin{figure}[h]
\epsfxsize=10cm \centerline{\epsfbox{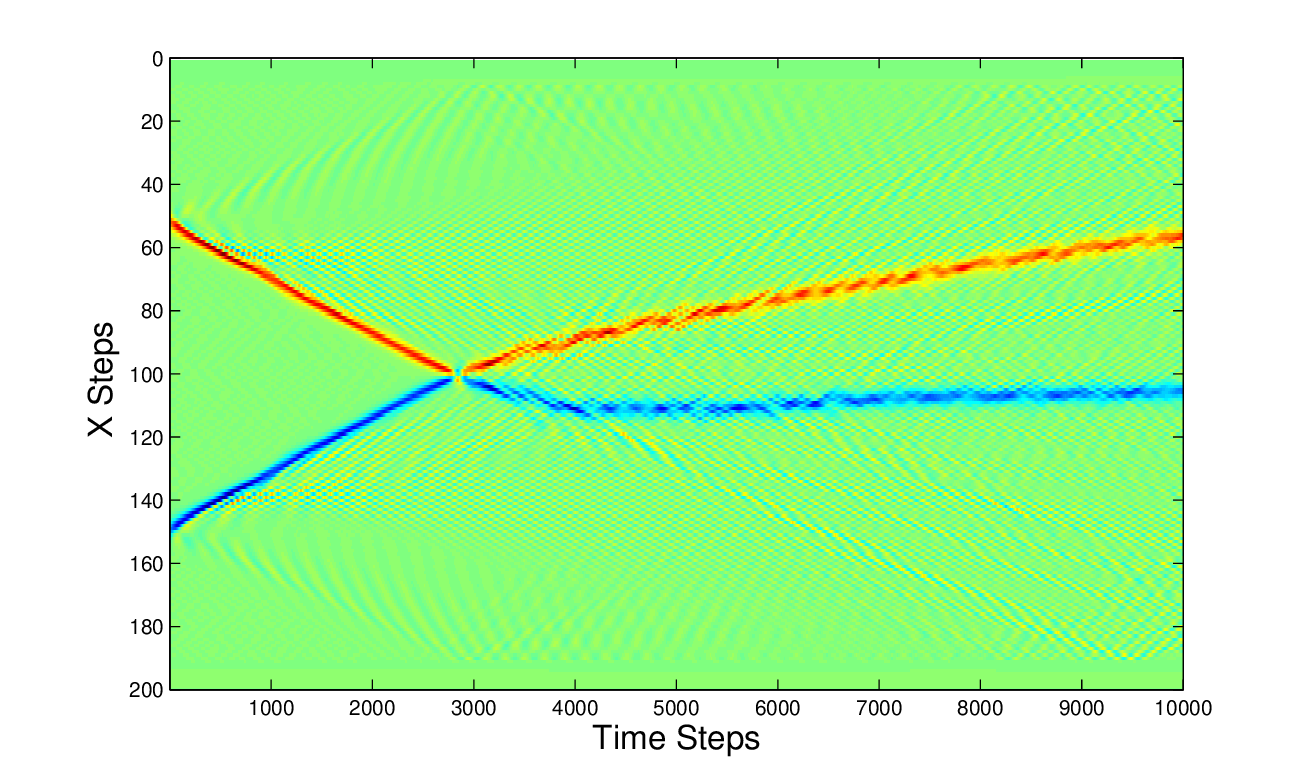}}
\caption{Kink-antikink interaction, leading to the excitation and
subsequent de-excitation and recoil of the kink and antikink
($v=0.7$). Red color indicates positive and blue indicates
negative topological charge.\label{decay}}
\end{figure}

Figure \ref{simple} shows simple scattering of a kink and
antikink. The velocity of each soliton (in units $c$) is 0.6 for
this process. Figures \ref{gamma}-\ref{decay} show examples of
some interesting interactions for the system considered in this
paper. In Figure \ref{gamma}, the pair annihilate each other into
a pair of neutral wave packets which leave the interaction area
with larger velocities. Figure \ref{molecule} shows the formation
of a bound pair emitting the residual energy in the form of lower
amplitude scalar waves.  In Figure \ref{decay}, the collision
leads to the excitation of each domain wall and the excited wall
relaxes into its lower energy state, emitting a neutral waves
within a short time. In these plots, positive topological charge
density is shown in red and negative charge in blue in order to
better illustrate the location and the fate of charged objects.

Quantum mechanically, the radiation emitted by the accelerating
kinks is in the form of scalar particles \cite{106,107}. In three
space dimensions, domain walls can undergo acceleration and
deformations due to  their own tension, except in the very special
cases of static solutions. The radiation emitted from deformed
domain walls has been calculated both analytically and numerically
\cite{182}. Radiation due to periodically deformed kinks has been
calculated analytically in \cite{110,140}.

\begin{figure}[h]
\epsfxsize=10cm \centerline{\epsfbox{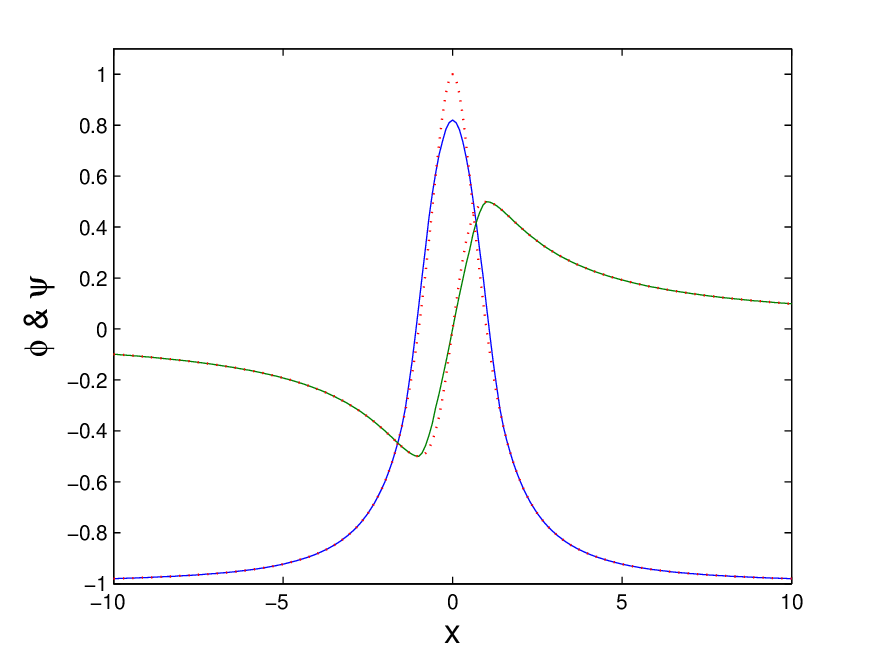}}
\caption{Initial guess functions (\ref{guess1}) and (\ref{guess2})
shown as dashed curves, together with the minimum energy
kink-antikink bound state solution (solid
lines).\label{solmolecule}}
\end{figure}

Let us examine the bound pair in more detail. If there is a stable
kink-antikink bound state, then one should be able to obtain it
via an energy-minimization procedure. To this end, we have
followed an energy-minimization algorithm, which produces a
minimum-energy solution, starting with a trial pair of functions
which satisfy the boundary conditions and the general functional
form of the soliton pair. The initial guess functions read
\begin{equation}\label{guess1}
\phi(x)=\frac{2}{1+x^2}-1,
\end{equation}
and
\begin{equation}\label{guess2}
\psi (x)=\frac{x}{1+x^2}.
\end{equation}
It is obvious that these trial functions have zero total
topological charges, comprising equal negative and positive
charges of the kink and antikink constituents. The initial guess,
together with the minimum energy solution are shown in Figure
\ref{solmolecule}. This minimum-energy bound state of the kink and
antikink closely conforms with the pair formed in the numerical
experiment shown in Figure \ref{molecule}.

\section{Can the domain walls get punctured?\label{punsec}}

\begin{figure}[h]
\epsfxsize=15cm \centerline{\epsfbox{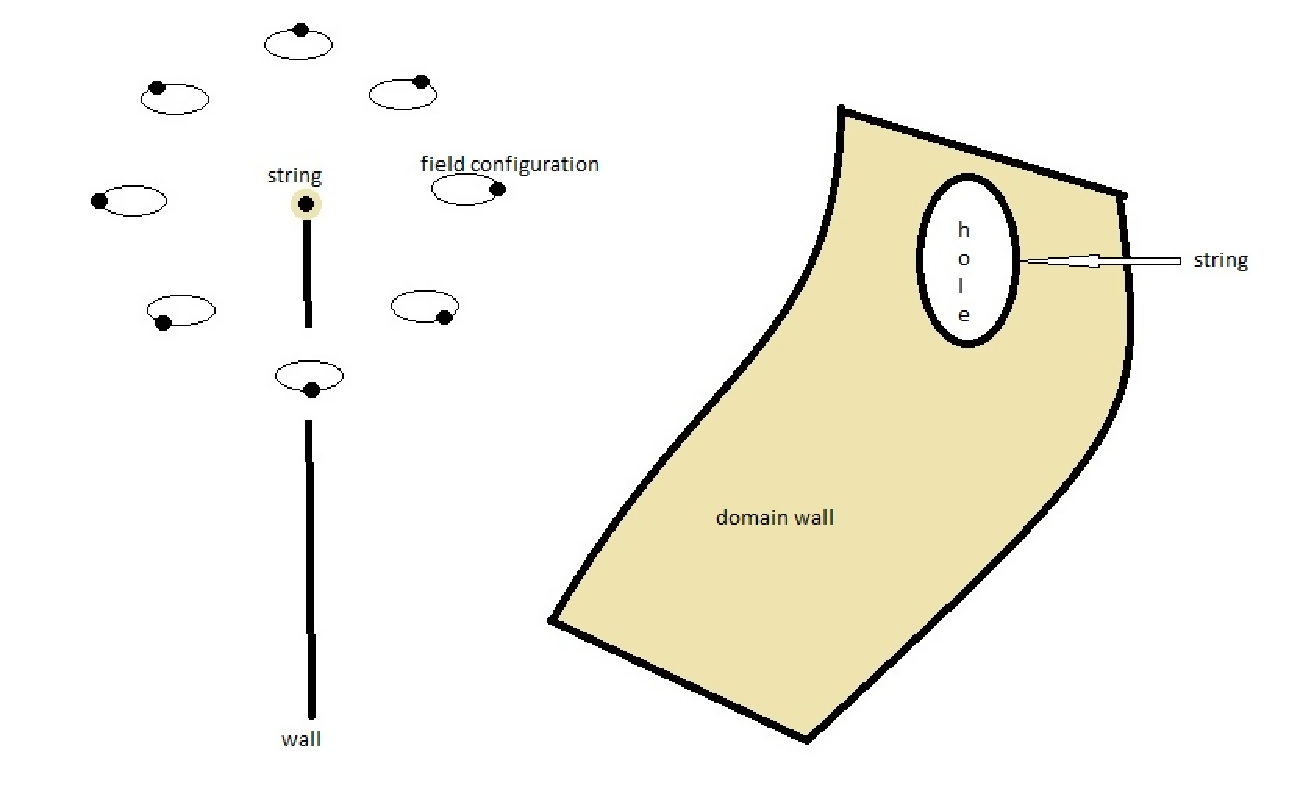}} \caption{Domain
wall punctured by a closed string. The small ellipses indicate the
field configuration on the ($\phi,\psi $) plane. It is assumed
that the $\psi$-field $Z_2$ symmetry is broken and $\psi>0$ vacuum
has a lower energy than the $\psi <0$ vacuum.\label{topfig}}
\end{figure}

Some domain walls can get punctured\cite{dom}. Here, we follow the
criteria presented in \cite{dom} to check if our domain walls can
get punctured, too. A punctured domain wall has a hole in it. The
boundary of this hole is formed by a closed string. The potential
considered in \cite{dom} is the following:
\begin{equation}
V(\Phi)=\frac{\lambda}{4}(|\Phi|^2-\eta^2)^2-\frac{\alpha\eta}{32}(\Phi+\Phi^*)^3,
\end{equation}
in which $0<\alpha<<\lambda$. It is seen that this potential
differs from (1) in the second term. The extrema of this potential
are located at $\psi=0$ and
\begin{equation}
\psi=\eta \left[ \frac{3\alpha +
\sqrt{9\alpha^2+64\lambda^2}}{8\lambda}\right], \ \ \chi=n\pi.
\end{equation}
Here, $\psi$ and $\chi$ are the module and phase of $\Phi$ (i.e.
$\Phi=\psi \exp(i\chi)$). A domain wall exists when we have two
disconnected vacua at boundaries (e.g. $\chi (-\infty)=0$ and
$\chi (+\infty)=2\pi$). Now, the path from $\chi= 0$ to $\chi =
2\pi$ can be contracted by lifting it over the top  of the
potential at $\psi=0$. In this way, a patch of the domain wall can
be bounded by a string and a hole can form\cite{kib} (see Figure
\ref{topfig}).

Now we turn to the potential (1) to see if it is topologically
possible to have the same situation. As mentioned in the
Introduction, we have two disconnected points at ($\phi=\pm 1$,
$\psi=0$) as true vacua. There is a potential barrier located at
$\phi=\psi=0$ (false vacuum) which separates these two points.
Therefore, it is topologically possible for the field to be
contracted by lifting it over the false vacuum. If the $Z_2$
symmetry in the direction of the $\psi$-field is broken by a term
like $\kappa \psi^n$, $n$=odd, then the punctured area will have a
lower energy density compared to the domain wall. The topology of
the field-3Dspace mapping is shown in Figure \ref{topfig}.

Whether the puncture tends to get larger and larger or likes pinch
off, is an interesting question which needs further investigation.

\section{conclusion}\label{sec4}
Kink-bearing systems show very interesting phenomena
\cite{0,1,2,3,4,5}. Domain walls are in fact mathematically the
same structures, extended in two more spatial dimensions.  A
complex scalar field with $U(1)$ symmetry is well known and worked
out thoroughly in field theory\cite{8}. When the global $U(1)$
symmetry is made local, it leads to the appearance of electric
charge and electromagnetic interactions. A system comprised of a
complex scalar field coupled to the U(1) gauge field with
spontaneously broken symmetry (the so-called abelian Higgs model)
is known to bear cosmic string solutions.  Motivated by these
interesting properties, we considered a double-real-field
Lagrangian with a $U(1)$-breaking term. We obtained static domain
wall solutions and showed that there are two degenerate pairs of
kinks and antikinks in the system, related to each other by the
symmetry operations $\phi\leftrightarrow \phi$ and
$\psi\leftrightarrow -\psi$. Several numerical experiments were
performed to explore what happens in the parallel collision of
domain walls at various relative velocities. It was observed that
different interesting phenomena may happen. Examples include
simple scattering, pair annihilation into neutral wave packets,
formation of soliton molecules (bound kink-antikink pairs), and
excitation-decay process. The soliton molecule formed in some
kink-antikink collisions approximately conforms with the solution
obtained via minimizing the energy of a pair of guess functions
adapted to the required topological charge and boundary
conditions.  In order to distinguish charged solitons from neutral
wave packets and follow the evolution of each charged soliton, we
preferred to plot charge densities rather than the fields or
energy densities which is more common in the literature.

Another observation to be pointed out is that in many examples,
the resulting dynamics is not symmetrical about the pair center of
mass. In other words, there is a left-right asymmetry which
constitutes yet another difference with other well-known
non-integrable systems. The model is also interesting in the sense
that in the case of vanishing coupling constant $\lambda=0$, it
reduces to a global U(1) system. For $0<\lambda<4$ two pairs of
degenerate kinks appear which are transformed to each other by a
parity transformation. For $\lambda>4$ the system essentially
reduces to the $\phi^4$ system.

Finally, we discussed the topological possibility for the domain
wall being punctured by a string loop. We argued that the form of
the potential allows the field to lift over the potential barrier
at $\phi=\psi=0$ and form a string at the boundary of the domain
wall, forming a hole bounded by a string.

\acknowledgements{N. Riazi acknowledges the support of Shahid
Beheshti University Research Council.}

\end{document}